\documentclass[preprint,amsmath,amssymb,amsfonts,aps,superscriptaddress,nofootinbib]{revtex4}
\usepackage{ifsym}
\usepackage{MnSymbol}
\usepackage[usenames,dvipsnames]{xcolor}
\usepackage{hyperref}
\def\be{\begin{equation}}
\def\ee{\end{equation}}
\def\ba{\begin{eqnarray}}
\def\ea{\end{eqnarray}}

\def\nn{\nonumber}

\def\a{\alpha}

\def\D{\Delta}

\def\L{\Lambda}
\def\l{\lambda}

\def\Th{\Theta}  

\def\cA{{\cal A}}

\def\[{\left[}
\def\]{\right]}
\def\({\left(}
\def\){\right)}

\def\<{\langle}
\def\>{\rangle}

\def\sun{{SU(N)}}  \def\son{SO(N)} \def\sp2n{Sp(2N)}

\def \bA{{\bar A}}

\def\2F1{\,_2{\rm F}_1}

\newcommand{\tr}{\text{Tr}}

\begin{document}
\title{Group constraint relations for five-point amplitudes in gauge theories with $\son$ and $\sp2n$ groups}

\author{Jia-Hui Huang}
\email{huangjh@m.scnu.edu.cn}
\affiliation{Guangdong Provincial Key Laboratory of Quantum Engineering and Quantum Materials,
School of Physics and Telecommunication Engineering,
South China Normal University, Guangzhou 510006,China}

\date{\today}

\begin{abstract}
In this paper,linear constraint relations among loop-order five-point  color-ordered amplitudes in $\son$ and $\sp2n$ gauge theories are derived with the group-theoretic method. These constrains are derived  up to four-loop order. It is found that in both theories, there are $n=6,22,34,44,50$ linear constraint relations at $L=0,1,2,3,4$ loop orders. Then the numbers of independent color-ordered five-point  amplitudes are respectively $n_{ind.}=6,12,22,34,50$ at each loop order.
\end{abstract}
\maketitle
\section{Introduction}
There has been a great advance in our understanding about perturbative amplitudes in gauge and gravity theories during last decade. One aspect of the advance is the revealing of many different kinds of relations between color-ordered gauge amplitudes and double-copy relations between gauge and gravity amplitudes.
For gauge amplitudes, one important discovery is the Bern-Carrasco-Johansson(BCJ) relation\cite{bcj2008}. These relations have been generally proved for tree level gauge amplitudes\cite{bdv2009,FHJ2011,cdf2011}. Subsequently, tree level BCJ relations have been proved in N=4 SYM \cite{Jia:2010nz},noncommutative $U(N)$ gauge theory \cite{hhj2011} and string theories\cite{Stieberger:2009hq,Mafra:2011kj,Lai:2016omd}. BCJ relations for gauge amplitudes with matter particles have been studied in \cite{Naculich:2014naa,delaCruz:2015dpa}. BCJ relations at loop orders have been explored in \cite{Bern:2010tq,cj2011,dl2013,Chester:2016ojq,He:2016mzd,He:2017spx}. Besides BCJ relations, tree level color-ordered gauge amplitudes also satisfy other linear constraint relations, such as $U(1)$ decoupling relations\cite{bg1987,mpx1988,bernk1991,FHJ2011,hhj2011} and Kleiss-Kuijf(KK) relations\cite{kk1989}.
$U(1)$ decoupling relations can be derived from the fact that $U(1)$ photon decouples from other gauge bosons and these relations can be extended to higher loop orders \cite{bernk1991}. The extension of KK relations to one- and two-loop has also been explored by generalized unitarity cut method \cite{ddm2000,bfd2002,cj2011,fjh2012,bi2012}.

In the past several years, a group-theoretic method has been proposed to derive linear constraints among loop order color-ordered amplitudes. One virtue of these constraints is that they hold for any helicity configuration of external particles. Some nontrivial all loop relations have been derived for four-point color-ordered amplitudes in $\sun$ gauge theories \cite{nacu2012}. Subsequently, the all loop relations for five- and six-point color-ordered amplitudes have also been derived \cite{enacu20121,enacu20122}. These loop-order constraint relations generalize known tree-level and one-loop  $\sun$ gauge amplitude relations.  Importantly, some of the nontrivial loop-order constraints are precisely those derived by direct calculations of loop amplitudes in $\sun$ gauge theories \cite{fjh2012,Tourkine:2016bak,Bern:1997nh}. Recently, this group-theoretic method has been applied to study color-ordered amplitudes in $\son$ and $\sp2n$ gauge theories \cite{Huang:2016iqf}. Linear constraint relations for four-point amplitudes in both kinds of theories are derived up to four-loop order.

In this paper, based on our previous work \cite{Huang:2016iqf}, we continue to use the group-theoretic method to derive linear constraints on five-point color-ordered amplitudes in $\son$ and $\sp2n$ gauge theories through four loop. All particles in the scattering amplitudes are adjoint particles.
It is found that, at $L=0,1,2,3,4$ loop orders, there are respectively $n=6,22,34,44,50$ group-theoretic constraint relations among five-point color-ordered amplitudes  in both  theories. The number of independent color-ordered amplitudes at each order for both theories is listed in Table 1.
\begin{table}
\caption{Loop orders vs. Numbers of independent amplitudes}
\centering
\begin{tabular}{|p{1.5cm}|p{0.8cm}|p{0.8cm}|p{0.8cm}|p{0.8cm}|p{0.8cm}|}
\hline
Loop&0 & 1 & 2 & 3 & 4\\\hline
Number &6 & 12 & 22 & 34 & 50 \\
\hline
\end{tabular}
\end{table}

The organization of this paper is as follows. In Section II, we present the trace basis and $L$-loop color decomposition of five-point amplitudes. In Section III, loop-level group-theoretic constraint relations among five-point color-ordered amplitudes in $\son$ theories are calculated up to four-loop order.
In section IV, we derive loop-order constraint relations among five-point amplitudes in $\sp2n$ gauge theories up to four-loop order. The last section is devoted to conclusion and discussion. Some  details in calculation are presented in the Appendix.

\section{Trace basis of five-point amplitudes}
It is known that gauge amplitudes can be decomposed into color parts and kinematic parts. The color parts can be expressed by products of structure constants $\{f^{abc}\}$ or by traces of generators in fundamental representation. The former decomposition can be called $f$-based decomposition and the latter can be called trace-based decomposition. Then, a full $n$-point gauge amplitude can be written as
 \ba\label{decom}
 \cA_n=\sum_\l c_\l a_\l=\sum_i t_i A_i,
 \ea
 where $\{c_\l\}$ is a color basis and $\{t_i\}$ is a trace basis. $a_\l$ is a kinematic factor corresponding to $c_\l$ and  $A_i$ is the color-ordered amplitude corresponding to $t_i$.

In the fundamental representation, the generators of $\son$ algebra are $N\times N$ antisymmetric, traceless matrices, which are denoted by $\{T^a\}$ ($a=1,2,...,N(N-1)/2$) \cite{Huang:2016iqf}. For five-point amplitudes, a trace basis consists of both single-trace and double-trace elements. A typical element in a trace basis is $\tr(T^{a_1}T^{a_2}T^{a_3}T^{a_4}T^{a_5})$, where $a_i$ is the color quantum number of the $i$-th particle. For simplicity, we use $\tr(12345)$ instead of $\tr(T^{a_1}T^{a_2}T^{a_3}T^{a_4}T^{a_5})$ in the following. The single-trace elements of a trace basis for $\son$ five-point amplitudes are
\ba\nonumber
T_1=\tr(12345)~~T_2=\tr(12354)~~T_3=\tr(12534)~~T_4=\tr(15234)~~T_5=\tr(12435) \\\nonumber
T_6=\tr(12453)~~T_7=\tr(12543)~~T_8=\tr(15243)~~T_9=\tr(14235)~~T_{10}=\tr(14253)\\
T_{11}=\tr(14523)~~T_{12}=\tr(15423).
 \ea
The double-trace elements of a trace basis for $\son$ five-point amplitudes are
\ba\nonumber
T_{13}=\tr(12)\tr(345)~T_{14}=\tr(13)\tr(245)~T_{15}=\tr(14)\tr(235)~T_{16}=\tr(15)\tr(234)\\\nn
T_{17}=\tr(23)\tr(145)~T_{18}=\tr(24)\tr(135)~T_{19}=\tr(25)\tr(134)~T_{20}=\tr(34)\tr(125)\\
T_{21}=\tr(35)\tr(124)~T_{22}=\tr(45)\tr(123).
 \ea
In the trace-based decomposition of loop amplitudes, terms are suppressed by different powers of $N$. So we extend the trace basis elements and decompose the amplitudes further in different powers of $N$\cite{Bern:1997nh,Huang:2016iqf}. Then, the color decomposition of $L$-loop $\son$ five-point full amplitude $\cA^L$ is
 \ba\label{socd}
 \cA^L=\sum_{m=0}^L\sum_{i=1}^{12} (N^{m}T_i) A_i^{(L,m)} +\sum_{n=0}^{L-1}\sum_{j=13}^{22} (N^{n}T_j) A_j^{(L,n)}.
 \ea
Finding linear constraints among color-ordered amplitudes $\{A_k^{(L,m)}\}$ is the aim of the paper and is studied with the group-theoretic method in Section III.

In the fundamental representation, the generators of $\sp2n$ algebra are $2N\times 2N$ antisymmetric, Hermitian matrices. The trace-based color decomposition  of five-point amplitudes in $\sp2n$ gauge theories is the same as $\son$ case. For simplicity, the trace basis elements of $\sp2n$ are denoted by the same symbols as the $\son$ case.  The color decomposition of $L$-loop $\sp2n$ five-point full amplitude $\bar{\cA}^L$ is
 \ba\label{spcd}
 \bar{\cA}^L=\sum_{m=0}^L\sum_{i=1}^{12} (N^{m}T_i) \bA_i^{(L,m)} +\sum_{n=0}^{L-1}\sum_{j=13}^{22} (N^{n}T_j) \bA_j^{(L,n)}.
 \ea
Linear constraints among $\sp2n$ color-ordered amplitudes $\{\bA_k^{(L,m)}\}$ are studied in Section IV.

\section{constraints on $\son$ five-point amplitudes}
In this section, we derive the linear constraints on $\son$ five-point amplitudes with group-theoretic method. This method is based on the fact that
the full gauge amplitudes can be decomposed by two ways: $f$-based  and trace-based decompositions. As in  eq.\eqref{decom}, the tree-level five-point full amplitude can be decomposed as
 \ba\label{treeeq}
 \cA^0=\sum_{i=1}^{12}T_i A_i^{(0,0)}=\sum_\l c_\l^0 a_\l^0.
 \ea
$c_\l^0$ is a tree-level color basis element, which is a product of several $\son$ structure constants. A $\son$ structure constant can always be expressed as a trace of its generators,
\ba\label{ftot}
 f^{abc}=-\frac{\text{i}}{2}\tr([T^a,T^b]T^c)=-\text{i}\tr(T^a T^b T^c).
 \ea
And with following two useful identities for the trace of $\son$ generators,
 \ba\label{iden}
 &&\tr(T^a A T^a B)=\tr(A)\tr(B)-(-1)^{n_B}(AB^r),\\
 &&\tr(T^a A)\tr(T^a B)=\tr(AB)-(-1)^{n_B}(AB^r),
 \ea
 a color basis can always be expressed as linear combination of a trace basis.

 At tree level,  there are six elements in a color basis $\{c_\l^0\}$, which can be chosen as $\{\text{i}^3 f^{12a}f^{a3b}f^{b45}, \text{all other permutations of (234)} \}$. The expansion matrix between this color basis and trace basis is $M^0$, defined as
 \ba
 c_\l^0=\sum_{k=1}^{12}M^0_{\l k}T_k,~(\l=1,2,\cdots,6).
 \ea
Combining with eq.\eqref{treeeq}, we can obtain the following relation
 \be
 A_k=\sum_\l a_\l M^0_{\l k}.
 \ee
 Then right null vectors of $M^0$ lead to linear constraints among color-ordered amplitudes $A_k (k=1,...,12)$.
The explicit form of $M^0$ is
\ba
M^0=2\left(
      \begin{array}{cccccccccccc}
        1& -1& 0 & 0 & 0 & -1 & 1 & 0 & 0 & 0 & 0 & 0 \\
        0 & -1 & 1 & 0 & 1 & -1 & 0 & 0 & 0 & 0 & 0 & 0 \\
        0 & 0 & 1 & -1 & 0 & 0 & 0 & 0 & 0 & -1 & 1 & 0 \\
        0 & 0 & 0 & 0 & 0 & -1 & 1& 0 & 0 & 0 & 1 & -1 \\
        0 & 0 & 0 & 0 & 0 & 0& 1 & -1 & 0 & -1 & 1 & 0 \\
        0 & -1 & 1 & 0 & 0 & 0 & 0 & 0 & 1 & -1 & 0 & 0 \\
      \end{array}
    \right).
\ea
The six tree-level null vectors of $M^0$, $r^{(0)}_p(p=1,...,6)$,  are listed as follows,
\ba\nn\label{treenullso}
(1,1,1,1,0_8),(0_4,1,1,1,1,0_4),(1,1,0_2,1,0_3,1,0_3),\\
(0_2,1,0_2,1,1,0_2,1,0_2),(0,-1,-1,0_3,-1,0_3,1,0),(-1,0_3,-1,-1,0_5,1),
\ea
where we use $(...,0_3,...)$ to denote $(...,0,0,0,...)$. They imply six constraints among twelve five-point color-ordered tree amplitudes. For example, the constraint equations implied by the first and last vectors are
 \ba
 A^{(0,0)}(1,2,3,4,5)+ A^{(0,0)}(1,2,3,5,4)+ A^{(0,0)}(1,2,5,3,4)+ A^{(0,0)}(1,5,2,3,4)=0\\
 -A^{(0,0)}(1,2,3,4,5)- A^{(0,0)}(1,2,4,3,5)- A^{(0,0)}(1,2,4,5,3)+ A^{(0,0)}(1,5,4,2,3)=0
 \ea
The first equation is obviously the same as tree-level dual Ward identity or KK-like relation in $\sun$ gauge theories. Using the reflection relation $A^{(0,0)}(1,5,4,2,3)=-A^{(0,0)}(1,3,2,4,5)$, we can see that the second equation is also a dual Ward identity or a KK-like relation. In fact, all six constraint relations are dual Ward identities or KK-like relation, which is the same as the $\sun$ case\cite{enacu20121}. There are six constraint relations among twelve tree-level five-point amplitudes, so the number of independent tree amplitudes is six.

\subsection{constraints among $\son$ one-loop five-point amplitudes}
In this section, we first explain the procedure to derive constraint relations on $L$-loop amplitudes. In order to find these constraints, we should find a complete (or overcomplete) $L$-loop color basis, expand elements of the color basis by $L$-loop trace basis and then find the null vectors of the expansion matrix. Each null vector leads to a constraint relation among $L$-loop color-ordered amplitudes.

It is easy to construct a $L$-loop trace basis. But constructing  a $L$-loop color basis is not easy. There is an assumption that all $(L+1)$-loop color factors can be obtained from $L$-loop color factors by attaching a rung between two of its external legs\cite{nacu2012}. This assumption can be checked explicitly at lower loop orders ($L\leqslant 4$) for $SU(N)$ and is thought to be correct at $L>4$ loop orders \cite{nacu2012}.
Here we adopt the same assumption and assume
that all possible elements of $(L+1)$-loop color basis can be obtained from $L$-loop elements by attaching external legs $(1,2),(1,3),(1,4),(1,5)$\cite{nacu2012,Huang:2016iqf}.

We consider the effect of this attaching process on the trace basis. Let
\ba
 T=\left(
  \begin{array}{c}
  T_1 \\
  T_2\\
  ...\\
  T_{12}\\
  \end{array}
  \right), ~\tilde{T}=\left(
                \begin{array}{c}
                  T_{13} \\
                  T_{14}\\
                  ...\\
                  T_{22}\\
                \end{array}
              \right).
  \ea
 After attaching a rung between two external legs, $T$ and $\tilde{T}$ transform as following,
 \ba\label{TraceTrans}
 T\rightarrow \(A,B,C\)\left(
                         \begin{array}{c}
                           N T \\
                           \tilde{T}\\
                           T\\
                         \end{array}
                       \right),~~~~~~
 \tilde{T}\rightarrow\(D,E,F\)\left(
                                \begin{array}{c}
                                  N \tilde{T} \\
                                  \tilde{T}\\
                                  T\\
                                \end{array}
                              \right),
 \ea
where the explicit forms of $A, B, C, D, E, F$ are given in the Appendix.

Suppose $\{c^{L}_\a\}$ is a complete color basis for $L$-loop five-point amplitudes and they can be expressed by a $L$-loop trace basis
$\{T^{L}_k, (k=1,2,\cdots,22L+12)\}=\{N^L T_i,N^{L-1} T_j,N^{L-1} T_i,\cdots, T_j, T_i, (i=1,2,...,12;j=13,...,22)\}$,
 \ba\label{Mjuzhen}
 c_\a^{L}=\sum_{k=1}^{22L+12} M_{\a k}^{L}T_k^{L}.
 \ea
According to eq.\eqref{TraceTrans}, the attaching procedure transforms $L$-loop trace basis $\{T_k^{L}\}$ to $(L+1)$-loop trace basis $\{T_l^{(L+1)}\}$,
 \ba
  T_k^{L}\rightarrow \sum_{l=1}^{22L+34} G_{kl}^{(L,L+1)}T_l^{(L+1)},
 \ea
where $G$ is a $(22L+12)\times(22L+34)$ transformation matrix. After the attaching procedure, $L$-loop color basis and trace basis are transformed to $(L+1)$-loop color basis and trace basis, and then we have
\ba
 c_\a^{(L+1)}=\sum_{l=1}^{22L+34} M_{\a l}^{(L+1)}T_l^{(L+1)}=\sum_{k=1}^{22L+12} M_{\a k}^{L}\sum_{l=1}^{22L+34}G_{kl}^{(L,L+1)}T_l^{(L+1)}.
 \ea
A $(L+1)$-loop right null vector $ r^{(L+1)}$  satisfies
 \ba
\sum_{l=1}^{22L+34} M_{\a l}^{(L+1)}r^{(L+1)}_l=\sum_{k=1}^{22L+12} M_{\a k}^{(L)}\sum_{l=1}^{22L+34}G_{kl}^{(L,L+1)}r^{(L+1)}_l=0,
 \ea
which means
 \ba\label{nuvere}
 G^{(L,L+1)}\cdot r^{(L+1)}=\textrm{linear combination of } \{r^{(L)}\}.
 \ea
This is the relation between $L$-loop and $(L+1)$-loop null vectors, which can be used to derive higher loop null vectors from lower loop ones.

Now we derive the one-loop null vectors from tree-level null vectors. The transformation matrix $G^{(0,1)}$ between trace bases of tree amplitudes and one-loop amplitudes is
 \ba
 G^{(0,1)}=(A,B,C).
 \ea
 Substituting $G^{(0,1)}$ and tree null vectors \eqref{treenullso} into eq.\eqref{nuvere}, we can obtain 22 one-loop null vectors, which can be written as a matrix $r^{(1)}=\(R^{(1)}_1, R^{(1)}_2\)$. Each column in matrices $R^{(1)}_1, R^{(1)}_2$ is one null vector.  $R^{(1)}_1, R^{(1)}_2$ are defined as
 \ba\label{R1so}
 R^{(1)}_1=\left(
         \begin{array}{c}
           m^{(1)}_1 \\
           \textbf{1}_{10\times 10} \\
           \textbf{0}_{12\times 10} \\
         \end{array}
       \right),
 R^{(1)}_2=\left(
         \begin{array}{c}
           m^{(1)}_2 \\
           \textbf{0}_{10\times 12} \\
           \textbf{1}_{12\times 12} \\
         \end{array}
       \right).
 \ea
The matrices $m^{(1)}_1, m^{(1)}_2$ are given by
 \ba\nn
m^{(1)}_1=\left(
        \begin{array}{cccccccccc}
          -1 & -1 & -1 & -1 & -1 & -1 & -1 & -1 & -1 & -1 \\
          1 & 1 & -1 & -1 & 1 & -1 & -1 & -1 & -1 & -1 \\
          -1 & 1 & 1& -1 & 1 & 1 & -1 & -1 & -1 & -1 \\
          -1 & -1 & -1 & -1 & 1 & 1 & -1 & 1 & -1 & -1 \\
          1 & -1 & -1 & 1 & -1 & -1 & 1 & -1 & -1 & -1 \\
          -1 & -1 & 1 & 1 & -1 & 1 & 1 & -1 & -1 & -1 \\
          1 & 1 & 1 & 1 & 1 & 1 & 1 & -1 & -1 & -1 \\
          1 & -1 & -1 & 1 & 1 & 1 & 1 & 1 & -1 & -1 \\
          1 & 1 & -1 & -1 & -1 & -1 & 1 & -1 & 1 & -1 \\
          -1 & 1 & 1 & -1 & -1 & 1 & 1 & -1 & 1 & -1 \\
          -1 & -1 & -1 & -1 & -1 & 1 & 1 & 1 & 1 & -1 \\
          1 & 1 & -1 &  -1& 1 & 1 & 1 & 1 & 1 & -1 \\
        \end{array}
      \right),
m^{(1)}_2=\left(
            \begin{array}{cccccccccccc}
              5 & 1 & 1 & 1 & 1 & 1 & -1& -1& 1 & -1& 1 & -1 \\
              1 & 5 & 1 & 1 & 1 & -1& 1 & -1& 1 & -1& -1& 1 \\
              1 & 1 & 5 & 1 & -1& 1 & 1 & -1& -1& 1 & -1& 1 \\
              1 & 1 & 1 & 5 & -1& 1 & -1& 1 & -1& -1& 1 & 1 \\
              1 & 1 & -1& -1& 5 & 1 & 1 & 1 & 1 & -1& 1 & -1 \\
              1 & -1& 1 & -1& 1 & 5 & 1 & 1 & -1& 1 & 1 & -1 \\
              -1& 1 & 1 & -1& 1 & 1 & 5 & 1 & -1& 1 & -1& 1 \\
              -1& 1 & -1& 1 & 1 & 1 & 1 & 5 & -1& -1& 1 & 1 \\
              1 & 1 & -1& -1& 1 & -1& 1 & -1& 5 & 1 & 1 & 1 \\
              1 & -1& 1 & -1& -1& 1 & 1 & -1& 1 & 5 & 1 & 1 \\
              1 & -1& -1& 1 & -1& 1 & -1& 1 & 1 & 1 & 5 & 1 \\
              -1& 1 & -1& 1 & -1& -1& 1 & 1 & 1 & 1 & 1 & 5 \\
            \end{array}
          \right).
 \ea
The number of one-loop five-point color-ordered amplitudes is 34 ($22L+12$). So the number of independent color-ordered amplitudes is 12. From eq.\eqref{socd}, the one-loop color decomposition of five-point amplitude is
\ba
\cA^1=\sum_{m=0}^1\sum_{i=1}^{12} (N^{m}T_i) A_i^{(1,m)} +\sum_{j=13}^{22} T_j A_j^{(1,0)}
\ea
The null vectors allow one to choose leading single-trace amplitudes, $A^{(1,1)_1}$ through $A^{(1,1)_{12}}$, as the one-loop independent amplitudes. And all other subleading single-trace or double-trace amplitudes can be written as linear combinations of them.
\subsection{constraints among $\son$ two-loop five-point amplitudes}
In this section we consider the constraints on two-loop amplitudes. The transformation matrix between one-loop and two-loop trace bases is
\ba
G^{(1,2)}= \left(
   \begin{array}{ccccc}
     A & B & C & 0 & 0 \\
     0 & D & 0 & E & F \\
     0 & 0 & A & B & C \\
   \end{array}
 \right).
 \ea
 Substituting $G^{(1,2)}$ and one-loop null vectors into eq.\eqref{nuvere}, we can obtain 34 two-loop null vectors. These null vectors can be written as
 $r^{(2)}=\(R^{(2)}_1,R^{(2)}_2,R^{(2)}_3 \)$. Each column in $R^{(2)}_i(i=1,2,3)$ is one null vector. $R^{(2)}_1,R^{(2)}_2,R^{(2)}_3 $ are defined by
 \ba\label{sonuve2}
 R^{(2)}_1=\left(
             \begin{array}{c}
               m^{(2)}_1 \\
               m^{(2)}_2 \\
               \textbf{1}_{12\times 12} \\
               \textbf{0}_{22\times 12}\\
             \end{array}
           \right),
 R^{(2)}_2=\left(
             \begin{array}{c}
                m^{(2)}_3 \\
               4*\textbf{1}_{10\times 10} \\
               \textbf{0}_{12\times 10} \\
               \textbf{1}_{10\times 10} \\
               \textbf{0}_{12\times 10}  \\
             \end{array}
           \right),
 R^{(2)}_3= \left(
              \begin{array}{c}
                m^{(2)}_4 \\
               m^{(2)}_5 \\
               \textbf{0}_{22\times 12} \\
               \textbf{1}_{12\times 12}\\
              \end{array}
            \right).
 \ea
 The explicit expressions of matrices $m^{(2)}_i$ are
 \ba
  m^{(2)}_1=\left(
              \begin{array}{cccccccccccc}
                7 &3/2& 0 &3/2&3/2 &0&-3/2 & 0 &0&3/2& 0&-3/2\\
              3/2 & 7 & 3/2 & 0 & 0 &-3/2& 0 &3/2&3/2& 0 &-3/2 & 0 \\
                0 &3/2& 7&3/2&-3/2&0 &3/2&0& 0& 3/2& 0 &-3/2\\
               3/2& 0 &3/2& 7 & 0&-3/2& 0&3/2& -3/2& 0 &3/2 & 0 \\
               3/2& 0 & -3/2& 0& 7& 3/2& 0& 3/2&3/2& 0 & -3/2 & 0 \\
                0 &-3/2& 0 & 3/2&3/2&7&3/2&0& 0 &3/2&0& -3/2\\
              -3/2& 0 &3/2& 0& 0&3/2&7&3/2& 3/2 & 0 &-3/2& 0 \\
                0 &-3/2& 0& 3/2&3/2&0&3/2&7& 0 &-3/2& 0 &3/2\\
                0 &3/2&0&-3/2&3/2&0&-3/2& 0& 7&3/2&0& 3/2\\
              -3/2& 0 & 3/2& 0& 0&3/2& 0&-3/2& 3/2 & 7 & 3/2 &0 \\
                0 &-3/2& 0 &3/2&3/2&0&-3/2&0& 0 & 3/2 &7 & 3/2 \\
              -3/2& 0 & 3/2&0& 0 &-3/2& 0& 3/2& 3/2 &0 & 3/2& 7 \\
              \end{array}
            \right),
 \ea
 \ba
  m^{(2)}_2=\frac{1}{4}\left(
                      \begin{array}{cccccccccccc}
                        1 & -1& 1 & -1& -1& 1 & -1& 1 & 1 & -1 & -1& 1 \\
                        -1& 1 & 1 & -1& -1& 1 & -1& 1 & 1 & -1 & 1 & -1 \\
                        -1& 1 & -1& 1 & -1& 1 & 1 & -1& 1 & -1 & 1 & -1 \\
                        1 & -1& -1& 1 & -1& 1 & 1 & -1& 1 & -1 & -1& 1 \\
                        1 & -1& 1 & -1& -1& -1& 1 & 1 & 1 & -1 & 1 & -1 \\
                        -1& -1& 1 & 1 & 1 & -1& 1 & -1& 1 & -1 & 1 & -1 \\
                        -1& -1& 1 & 1 & 1 & 1 & -1& -1& 1 & -1 & -1& 1 \\
                        1 & -1& 1 & -1& 1 & -1& 1 & -1& -1& -1 & 1 & 1 \\
                        -1& 1 & 1 & -1& 1 & 1 & -1& -1& -1& -1 & 1 & 1 \\
                        1 & 1 & -1& -1& -1& 1 & 1 & -1& -1& -1 & 1 & 1 \\
                      \end{array}
                    \right),
 \ea
 \ba
  m^{(2)}_3=2\left(
               \begin{array}{cccccccccc}
               -1  & -1& -1& -1& -1& -1& -1& -1& -1& -1 \\
                 1 & 1 & -1& -1& 1 & -1& -1& -1& -1& -1 \\
                 -1& 1 & 1 & -1& 1 & 1 & -1& -1& -1& -1 \\
                 -1& -1& -1& -1& 1 & 1 & -1& 1 & -1& -1 \\
                 1 & -1& -1& 1 & -1& -1& 1 & -1& -1& -1 \\
                 -1& -1& 1 & 1 & -1& 1 & 1 & -1& -1& -1 \\
                 1 & 1 & 1 & 1 & 1 & 1 & 1 & -1& -1& -1 \\
                 1 & -1& -1& 1 & 1 & 1 & 1 & 1 & -1& -1 \\
                 1 & 1 & -1& -1& -1& -1& 1 & -1& 1 & -1 \\
                 -1& 1 & 1 & -1& -1& 1 & 1 & -1& 1 & -1 \\
                 -1& -1& -1& -1& -1& 1 & 1 & 1 & 1 & -1 \\
                 1 & 1 & -1& -1& 1 & 1 & 1 & 1 & 1 & -1 \\
               \end{array}
             \right),
 \ea
 \ba
  m^{(2)}_4=\left(
              \begin{array}{cccccccccccc}
                -10&-6& 6 & -6& -6& 6 & 6 & -6& 6 &-18& 6 & 6 \\
                -6&-10&-6 & 6 & 6 & 6 & 6 &-18& -6& -6& 6 & 6\\
                6 & -6&-10& -6& 6 & 6 & -6& -6& -6& -6& -6& 18\\
                -6& 6 & -6&-10& -6& 18& -6& -6& 6 & -6& -6& 6 \\
                -6& 6 &6  & -6&-10& -6& 6 & -6& -6& -6& 18& -6\\
                6 & 6 & 6 &-18& -6&-10& -6& 6 & -6& -6& 6 & 6 \\
                6 & 6 & -6& -6& 6 & -6&-10& -6&-18& 6 & 6 & 6 \\
                -6& 18& -6& -6& -6& 6 & -6&-10& -6& 6 & 6 & -6\\
                6 & -6& -6& 6 & -6& -6& 18& -6&-10& -6& 6 & -6\\
                18& -6& -6& -6& -6& -6& 6 & 6 & -6&-10& -6& 6 \\
                6 & 6 & -6& -6&-18& 6 & 6 & 6 & 6 & -6&-10& -6\\
                6 & 6 &-18& 6 & -6& 6 & 6 & -6& -6& 6 & -6& -10\\
              \end{array}
            \right),
 \ea
 \ba
  m^{(2)}_5=2\left(
               \begin{array}{cccccccccccc}
                 -1& 1 & -1& 1 & 1 & -1& 1 & -1& -1& 1 & 1 & -1 \\
                 1 & -1& -1& 1 & 1 & -1& 1 & -1& -1& 1 & -1& 1 \\
                 1 & -1& 1 & -1& 1 & -1& -1& 1 & -1& 1 & -1& 1 \\
                 -1& 1 & 1 & -1& 1 & -1& -1& 1 & -1& 1 & 1 & -1 \\
                 -1& 1 & -1& 1 & 1 & 1 & -1& -1& -1& 1 & -1& 1 \\
                 1 & 1 & -1& -1& -1& 1 & -1& 1 & -1& 1 & -1& 1 \\
                 1 & 1 & -1& -1& -1& -1& 1 & 1 & -1& 1 & 1 & -1\\
                 -1& 1 & -1& 1 & -1& 1 & -1& 1 & 1 & 1 & -1& -1\\
                 1 & -1& -1& 1 & -1& -1& 1 & 1 & 1 & 1 & -1& -1\\
                 -1& -1& 1 & 1 & 1 & -1& -1& 1 & 1 & 1 & -1& -1 \\
               \end{array}
             \right).
 \ea
These 34 two-loop null vectors imply 34 linear constraints among color-ordered five-point amplitudes. At two-loop order, the number of color-ordered amplitudes is 56($22L+12$).So the number of independent amplitudes is 22. From eq.\eqref{sonuve2}, we observe that we can choose the amplitudes, which correspond to leading single-trace (12 terms) and leading double-trace (10 terms) basis elements, as independent color-ordered two-loop five-point amplitudes. All other amplitudes can be expressed as linear combinations of them.
\subsection{constraints among $\son$ three,four-loop amplitudes}
In this section, we discuss the constraints among three- and four-loop five point amplitudes. The transformation matrix $G^{(2,3)}$ between two-loop null vectors and three-loop null vectors is
 \ba
 G^{(2,3)}=\left(
             \begin{array}{ccccccc}
               A & B & C & 0 & 0 & 0 & 0 \\
               0 & D & 0 & E & F & 0 & 0 \\
               0 & 0 & A & B & C & 0 & 0 \\
               0 & 0 & 0 & D & 0 & E & F \\
               0 & 0 & 0 & 0 & A & B & C \\
             \end{array}
           \right).
 \ea
 By solving eq.\eqref{nuvere}, we can obtain 44 three-loop null vectors. There are 78($22L+12$) three-loop color-ordered amplitudes, so the number of independent three-loop amplitudes are 34. We do not list the null vectors explicitly.

 The transformation matrix $G^{(3,4)}$ between two-loop null vectors and three-loop null vectors is
 \ba
 G^{(3,4)}=\left(
             \begin{array}{ccccccccc}
               A& B & C & 0 & 0 & 0& 0& 0& 0\\
               0 & D& 0 & E & F& 0 & 0 & 0 & 0 \\
               0 & 0 & A& B & C& 0& 0 &0 & 0 \\
               0 & 0& 0& D & 0 & E & F & 0 & 0 \\
               0 & 0 & 0 & 0 & A & B & C & 0 & 0 \\
               0 & 0 & 0 & 0 & 0 & D & 0 & E & F \\
               0 & 0 & 0 & 0 & 0 & 0 & A & B & C \\
             \end{array}
           \right).
 \ea
 By solving eq.\eqref{nuvere}, we can obtain 50 four-loop null vectors. There are 100($22L+12$) four-loop color-ordered amplitudes, so the number of independent four-loop amplitudes are 50. We do not list the null vectors explicitly.

\section{constraints on $\sp2n$ five-point amplitudes}
In this section, we use the same procedure to derive group-theoretic constraints among color-ordered $\sp2n$ five-point amplitudes. At tree level, six elements in a $\sp2n$ color basis can  chosen as $\{\text{i}^3 F^{12a}F^{a3b}F^{b45}, \text{all other permutations of (234)} \}$($F^{abc}$ is a $\sp2n$ structure constant). The expansion matrix $M^0$  between this tree-level color basis and trace basis is the same as $\son$ case,
\ba
M^0=2\left(
      \begin{array}{cccccccccccc}
        1& -1& 0 & 0 & 0 & -1 & 1 & 0 & 0 & 0 & 0 & 0 \\
        0 & -1 & 1 & 0 & 1 & -1 & 0 & 0 & 0 & 0 & 0 & 0 \\
        0 & 0 & 1 & -1 & 0 & 0 & 0 & 0 & 0 & -1 & 1 & 0 \\
        0 & 0 & 0 & 0 & 0 & -1 & 1& 0 & 0 & 0 & 1 & -1 \\
        0 & 0 & 0 & 0 & 0 & 0& 1 & -1 & 0 & -1 & 1 & 0 \\
        0 & -1 & 1 & 0 & 0 & 0 & 0 & 0 & 1 & -1 & 0 & 0 \\
      \end{array}
    \right).
\ea
The six tree-level null vectors of $M^0$ is also the same as $\son$ case,
\ba\nn\label{treenullsp}
(1,1,1,1,0_8),(0_4,1,1,1,1,0_4),(1,1,0_2,1,0_3,1,0_3),\\
(0_2,1,0_2,1,1,0_2,1,0_2),(0,-1,-1,0_3,-1,0_3,1,0),(-1,0_3,-1,-1,0_5,1).
\ea
These 6 null vectors imply 6 linear constraint relations among 12 tree-level color-ordered $\sp2n$ five-point amplitudes. Therefore, the number of independent amplitudes is 6 at tree-level.

To find $(L+1)$-loop null vectors from $L$-loop null vectors, the core object is the transformation matrix $\bar{G}^{(L,L+1)}$ between $\sp2n$ $L$-loop trace basis and  $(L+1)$-loop trace basis due to the attaching procedure \cite{Huang:2016iqf}. In deriving the $\sp2n$ transformation matrix $\bar{G}^{(L,L+1)}$, the following two identities are important,
 \ba\label{sp2niden}\nonumber
 &&\tr(T^a M T^a N)=\tr(M)\tr(N)+(-1)^{n_N}(MN^r),\\
 &&\tr(T^a M)\tr(T^a N)=\tr(MN)-(-1)^{n_N}(MN^r).
 \ea

The transformation matrix $\bar{G}^{(0,1)}$ between trace bases of tree amplitudes and one-loop amplitudes is
 \ba
 \bar{G}^{(0,1)}=(\bar{A},\bar{B},\bar{C}).
 \ea
 The explicit forms of $\bar{A},\bar{B},\bar{C}$ are given in the appendix.
 Solving recursive equation \eqref{nuvere}, we can obtain 22 one-loop null vectors, which can be written as a matrix $\bar{r}^{(1)}=\(\bar{R}^{(1)}_1, \bar{R}^{(1)}_2\)$. Each column in matrices $\bar{R}^{(1)}_1, \bar{R}^{(1)}_2$ is one null vector.  $\bar{R}^{(1)}_1, \bar{R}^{(1)}_2$ are defined as
 \ba
 \bar{R}^{(1)}_1=\left(
         \begin{array}{c}
           \bar{m}^{(1)}_1 \\
           \textbf{1}_{10\times 10} \\
           \textbf{0}_{12\times 10} \\
         \end{array}
       \right),
 \bar{R}^{(1)}_2=\left(
         \begin{array}{c}
           \bar{m}^{(1)}_2 \\
           \textbf{0}_{10\times 12} \\
           \textbf{1}_{12\times 12} \\
         \end{array}
       \right).
 \ea
 These null vectors have similar structures with the one-loop $\son$ null vectors. Compared with eq.\eqref{R1so}, we have following relations between $\sp2n$ and $\son$ null vectors,
 \ba
\bar{m}^{(1)}_1=\frac{1}{2} m^{(1)}_1, ~~\bar{m}^{(1)}_2=-\frac{1}{2} m^{(1)}_2.
 \ea
These 22 null vectors imply 22 linear constraint relations among 34($22L+12$) one-loop color-ordered $\sp2n$ five-point amplitudes. Therefore, the number of independent  one-loop amplitudes is 12.

The transformation matrix $\bar{G}^{(1,2)}$ between trace bases of $\sp2n$ one-loop amplitudes and two-loop amplitudes is
\ba
\bar{G}^{(1,2)}= \left(
   \begin{array}{ccccc}
     \bar{A} & \bar{B} & \bar{C} & 0 & 0 \\
     0 & \bar{D} & 0 & \bar{E} & \bar{F} \\
     0 & 0 & \bar{A} & \bar{B} & \bar{C} \\
   \end{array}
 \right).
 \ea
 The explicit forms of the block matrices, $\bar{A},\cdots,\bar{F}$, are given in the appendix. Substituting $\bar{G}^{(1,2)}$ and one-loop null vectors into eq.\eqref{nuvere}, we can obtain 34 two-loop null vectors. These null vectors can be written as
 $\bar{r}^{(2)}=\(\bar{R}^{(2)}_1,\bar{R}^{(2)}_2,\bar{R}^{(2)}_3 \)$. Each column in $\bar{R}^{(2)}_i(i=1,2,3)$ is one null vector. $\bar{R}^{(2)}_1,\bar{R}^{(2)}_2,\bar{R}^{(2)}_3 $ are defined by
 \ba\label{sonuve2}
 \bar{R}^{(2)}_1=\left(
             \begin{array}{c}
               \bar{m}^{(2)}_1 \\
               \bar{m}^{(2)}_2 \\
               \textbf{1}_{12\times 12} \\
               \textbf{0}_{22\times 12}\\
             \end{array}
           \right),
 \bar{R}^{(2)}_2=\left(
             \begin{array}{c}
                \bar{m}^{(2)}_3 \\
               -2*\textbf{1}_{10\times 10} \\
               \textbf{0}_{12\times 10} \\
               \textbf{1}_{10\times 10} \\
               \textbf{0}_{12\times 10}  \\
             \end{array}
           \right),
 \bar{R}^{(2)}_3= \left(
              \begin{array}{c}
                \bar{m}^{(2)}_4 \\
               \bar{m}^{(2)}_5 \\
               \textbf{0}_{22\times 12} \\
               \textbf{1}_{12\times 12}\\
              \end{array}
            \right).
 \ea
These 34 two-loop null vectors are also similar to the $\son$ case. The block matrices in the $\sp2n$ null vectors satisfy following relations with the $\son$ case \eqref{sonuve2},
 \ba
 \bar{m}^{(2)}_1=-\frac{1}{2} m^{(2)}_1,~\bar{m}^{(2)}_2=- m^{(2)}_2, ~\bar{m}^{(2)}_3=-\frac{1}{4} m^{(2)}_3,~ \bar{m}^{(2)}_4=\frac{1}{4} m^{(2)}_4,~\bar{m}^{(2)}_5=\frac{1}{2} m^{(2)}_5.
 \ea
These 34 null vectors imply 34 linear constraint relations among 56 ($22L+12$) two-loop color-ordered $\sp2n$ five-point amplitudes. Therefore, there are 22 independent  two-loop five-point amplitudes.

The derivation of three- and four-loop constraints is similar with $\son$ case. The transformation matrices have the similar forms in both kinds of theories. We can obtain 44 null vectors at three-loop order for $\sp2n$ five-point amplitudes and the number of independent color-ordered three-loop amplitudes is 34. At four-loop order, there are 50 null vectors and the number of independent color-ordered amplitudes is 50.

\section{conclusion and discussion}
In this paper, we use the group-theoretic method to derive linear constraint relations among loop-order five-point  color-ordered amplitudes in $\son$ and $\sp2n$ gauge theories.  These constrains are derived  up to four-loop order due to the large dimension of null vectors at higher-loop orders and the computational accuracy problem, which is similar to four-point amplitude case\cite{Huang:2016iqf}. The block matrices, $A,\cdots,F$, of the transformation matrices $G^{(L,L+1)}$ in $\son$ theory  have similar structures with $(\bar{A},\cdots,\bar{F})$ in $\sp2n$ theory and the tree-level null vectors in both theories are same. So the loop-order null vectors in both theories have similar forms and values. In both theories, there are $n=6,22,34,44,50$ linear constraint relations at $L=0,1,2,3,4$ loop orders. Then the numbers of independent amplitudes are $n_{ind.}=6,12,22,34,50$ at each loop order respectively. We can choose the most-leading $n_{ind.}$ color-ordered amplitudes as independent amplitudes at each loop order and all other subleading  amplitudes can be expressed as linear combinations of them.

Group-theoretic  constraint relations among loop-order five-point $\sun$ gauge amplitudes is discussed in \cite{enacu20121}. It is shown that there are $n=6,10,12,10,12$ linear constraint relations at $L=0,1,2,3,4$ loop orders and the numbers of  independent amplitudes are $n_{ind.}=6,12,22,34,44$ at each loop order. Compared with the $\son$ and $\sp2n$ cases, the numbers of  independent amplitudes are the same up to 3-loop order, but different at four-loop order.
This is an unexpected result because the number of independent of amplitudes is equal to the number of independent color-basis elements at each loop order and
the number of  independent color-basis elements is expected to have nothing to do with details of the group algebras. This problem is related with the number of independent color basis elements and needs further study.

{\textbf{Acknowledgements:\\}}
This work is supported by the Natural Science Foundation of Guangdong Province (No.2016A030313444).

\appendix*
\section{transformation matrix}
In this appendix, we give the explicit forms of the block matrices in transformation matrix $G^{(L,L+1)}$ in both theories. Let quantity $e_{1i}$ take one when we attach legs $(1,i)$ and otherwise take zero. Define $\L_{ij}=e_{1i}+e_{1j}$ , $\D_{ij}=e_{1i}-e_{1j}$ and  $\Th_{ij,kl}=3\L_{ij}-\L_{kl}$.
For $\son$ case, the block matrices $A,\cdots,F$ are of the following forms
 \be\footnotesize
 A=\left(
     \begin{array}{cccccccccccc}
       -\L_{25} & 0 & 0 & 0 & 0 & 0 & 0 & 0 & 0 & 0 & 0 & 0 \\
       0 & -\L_{24} & 0 & 0 & 0 & 0 & 0 & 0 & 0 & 0 & 0 & 0 \\
       0 & 0 & -\L_{24} & 0 & 0 & 0 & 0 & 0 & 0 & 0 & 0 & 0 \\
       0 & 0 & 0 & -\L_{45} & 0 & 0 & 0 & 0 & 0 & 0 & 0 & 0 \\
       0 & 0 & 0 & 0 & -\L_{25} & 0 & 0 & 0 & 0 & 0 & 0 & 0 \\
       0 & 0 & 0 & 0 & 0 & -\L_{23} & 0 & 0 & 0 & 0 & 0 & 0 \\
       0 & 0 & 0 & 0 & 0 & 0 & -\L_{23} & 0 & 0 & 0 & 0 & 0 \\
       0 & 0 & 0 & 0 & 0 & 0 & 0 & -\L_{35} & 0 & 0 & 0 & 0 \\
       0 & 0 & 0 & 0 & 0 & 0 & 0 & 0 & -\L_{45} & 0 & 0 & 0 \\
       0 & 0 & 0 & 0 & 0 & 0 & 0 & 0 & 0 & -\L_{34} & 0 & 0 \\
       0 & 0 & 0 & 0 & 0 & 0 & 0 & 0 & 0 & 0 & -\L_{34} & 0 \\
       0 & 0 & 0 & 0 & 0 & 0 & 0 & 0 & 0 & 0 & 0 & -\L_{35} \\
     \end{array}
   \right),
 \ee

 \be\footnotesize
  B=\left(
      \begin{array}{cccccccccc}
        \D_{32} & 0 & 0 & \D_{45} & \D_{34} & 0 & 0 & 0 & 0 & \D_{43} \\
        \D_{23} & 0 & \D_{54} & 0 & \D_{53} & 0 & 0 & 0 & 0 & \D_{53}\\
       \D_{52} & 0 & \D_{43} & 0 & 0 & 0 & \D_{53} & \D_{35} &0 & 0\\
        0 & 0 & \D_{34} & \D_{25} & 0 & 0 & \D_{23} & \D_{23} & 0 & 0 \\
       \D_{24} & 0 & 0 & \D_{53} & 0 & \D_{43} & 0 & 0 & \D_{34} & 0 \\
       \D_{42} & \D_{53} & 0 & 0 & 0 & \D_{54} & 0 & 0 & \D_{54} & 0 \\
        \D_{25} & \D_{34} & 0 & 0 & 0 & 0 &\D_{45} & \D_{45} &0 & 0 \\
        0 & \D_{43} & 0 & \D_{52} & 0 & 0 &\D_{42} & \D_{24} & 0 & 0 \\
        0 & 0 & \D_{24} & \D_{35} & 0 & \D_{23} & 0 & 0 & \D_{23} & 0 \\
        0 & \D_{35} & \D_{42} & 0 & 0 & \D_{52} & 0 & 0 & \D_{25} & 0\\
        0 & \D_{23} & \D_{54} & 0 & \D_{25} & 0 & 0 & 0 & 0 & \D_{52} \\
        0 & \D_{32} & 0 & \D_{45} & \D_{42} & 0 & 0 & 0 & 0 & \D_{42} \\
      \end{array}
    \right),
 \ee
 \be\footnotesize
 C=\left(
     \begin{array}{cccccccccccc}
       \Th_{25,34} & \D_{34} & 0 & \D_{54} & 0 & 0 &\D_{32} & 0 & 0 & 0 & 0 & \D_{34} \\
      \D_{35} &\Th_{24,35} & 0 & 0 & 0 & \D_{32} & 0 & 0 & \D_{45} & 0 & \D_{35} & 0 \\
       0 & 0 & \Th_{24,35} & \D_{35} & \D_{52} & 0 & \D_{53} & 0 & 0 &\D_{43} & 0 & 0 \\
       \D_{52} & 0 & \D_{32} & \Th_{45,23} & 0 & 0 & 0 & \D_{23} & 0 & 0 & \D_{43} & 0 \\
       0 & 0 & \D_{42} & 0 & \Th_{25,34}& \D_{43} & 0 & \D_{53} & \D_{34} & 0 & 0 & 0 \\
       0 & \D_{42} & 0 & 0 &\D_{45} &\Th_{23,45} & 0 & 0 & 0 & \D_{54} & 0 & \D_{53} \\
       \D_{52} & 0 & \D_{54} & 0 & 0 & 0 & \Th_{23,45} & \D_{45} & 0 & 0 & \D_{43} & 0 \\
       0 & 0 & 0 & \D_{24} & \D_{52} & 0 & \D_{42} & \Th_{35,24} & 0 & \D_{43} & 0 & 0 \\
       0 & \D_{42} & 0 & 0 &\D_{32} & 0 & 0 & 0 & \Th_{45,23} & \D_{23} & 0 & \D_{53} \\
       0 & 0 & \D_{42} & 0 & 0 & \D_{52} & 0 & \D_{53} & \D_{25}& \Th_{34,25} & 0 & 0 \\
       0 & \D_{25} & 0 & \D_{45} & 0 & 0 &\D_{23} & 0 & 0 & 0 & \Th_{34,25} & \D_{25}\\
       \D_{24} & 0 & 0 & 0 & 0 &\D_{23}& 0 & 0 &\D_{54} & 0 & \D_{24} & \Th_{35,24} \\
     \end{array}
   \right),
 \ee

 \be\footnotesize
 D=\left(
     \begin{array}{cccccccccc}
       -2e_{12} & 0 & 0 & 0 & 0 & 0 & 0 & 0&0 & 0 \\
       0 & -2e_{13} & 0 & 0 & 0 & 0 & 0 & 0 & 0 & 0 \\
       0 & 0 & -2e_{14} & 0& 0 & 0 & 0 & 0 & 0 & 0 \\
       0 & 0 & 0 & -2e_{15} & 0 & 0 & 0 & 0 & 0 & 0 \\
       0 & 0 & 0 & 0 & -\L_{54} & 0 & 0 & 0 & 0 & 0 \\
       0 & 0 & 0 & 0 & 0 & -\L_{53} & 0 & 0 & 0 & 0 \\
       0 & 0 & 0 & 0 & 0 & 0 & -e_{14}-e_{13} & 0 & 0 & 0 \\
       0 & 0 & 0 & 0 & 0 & 0 & 0& -e_{15}-e_{12} & 0 & 0 \\
       0 & 0 & 0 & 0 & 0 & 0 & 0 & 0 & -e_{14}-e_{12}& 0 \\
       0 & 0 & 0 & 0 & 0 & 0 & 0 & 0& 0 & -e_{13}-e_{12} \\
     \end{array}
   \right),
 \ee
 \be\footnotesize
 E=-2*D,
 \ee
 \be\footnotesize
 F=2\left(
     \begin{array}{cccccccccccc}
       \D_{35} & \D_{43} &\D_{54} & 0 & \D_{54} & \D_{43} & \D_{35} &0 & 0& 0& 0 & 0 \\
       0& 0 & 0 & 0 & 0 &\D_{52} & \D_{24} & \D_{45} & 0 & \D_{45}& \D_{24} & \D_{52} \\
       0 & \D_{52} & \D_{23} & \D_{35} & 0 & 0 & 0 & 0 & \D_{25} & \D_{32} & \D_{53} & 0 \\
      \D_{42} & 0 & 0 & \D_{24} & \D_{23} & 0 & 0 & \D_{32} & \D_{34} & 0 & 0 & \D_{43}\\
       \D_{32}& \D_{23} & 0 & 0 & 0 & 0 & 0 & 0 & 0 & 0 & \D_{23} & \D_{32} \\
       0 & 0& 0 & 0 & \D_{42} & \D_{24} & 0 & 0 &\D_{24} & \D_{42} & 0 & 0 \\
       0 & 0 & \D_{52} & \D_{25} & 0 & 0 &\D_{25}& \D_{52} & 0 & 0 & 0 & 0 \\
       0 & 0 & \D_{34} & \D_{43} & 0 & 0 & \D_{43} & \D_{34} & 0 & 0 & 0 & 0 \\
       0 & 0 & 0 & 0 & \D_{35} & \D_{53} & 0 & 0 & \D_{53} & \D_{35} & 0 & 0 \\
       \D_{45} & \D_{54} & 0 & 0 & 0& 0 & 0 & 0 & 0 & 0 & \D_{54} & \D_{45} \\
     \end{array}
   \right).
 \ee

For $\sp2n$, the block matrices $\bar{A},\cdots,\bar{F}$ are related with those of $\son$ and they satisfy the following equations
 \be\footnotesize
 \bar{A}=2A,~~\bar{B}=B,~~\bar{C}=-C,~~\bar{D}=2D,~~\bar{E}=-E,~~\bar{F}=F.
 \ee

\end{document}